\documentclass[12pt]{iopart}
\pdfoutput=1
\usepackage{color}
\usepackage[pdftex]{graphicx}
\usepackage[pdftex]{hyperref}
\hypersetup{colorlinks=true,citecolor=blue,linkcolor=red,urlcolor=blue}
\usepackage[all]{hypcap}
\usepackage[normalem]{ulem}
\usepackage{etoolbox}

\begin{document}

\title{Lifted graphene nanoribbons on gold: from atomic smooth sliding to multiple stick-slip regimes}

\author{L Gigli,$^1$ N Manini,$^2$ E Tosatti,$^{1,3,4}$ R Guerra,$^{2,5}$ A Vanossi$^{4,1}$}
\address{$^1$International School for Advanced Studies (SISSA), Via Bonomea 265, 34136 Trieste, Italy}
\address{$^2$Dipartimento di Fisica, Universit\`a degli Studi di Milano, Via Celoria 16, 20133 Milano, Italy}
\address{$^3$The Abdus Salam International Centre for Theoretical Physics (ICTP), Strada Costiera 11, 34151 Trieste, Italy}
\address{$^4$CNR-IOM Democritos National Simulation Center, Via Bonomea 265, 34136 Trieste, Italy}
\address{$^5$Center for Complexity and Biosystems, University of Milan, 20133 Milan, Italy}

\begin{abstract}
Graphene nanoribbons (GNRs) physisorbed on a Au(111) surface can be picked
up, lifted at one end, and made slide by means of the tip of an atomic-force microscope. 
The dynamical transition from smooth sliding to multiple stick-slip 
regimes, the pushing/pulling force asymmetry, the presence of pinning, and its
origin are real frictional processes in a nutshell, in need of a
theoretical description. 
To this purpose, we conduct classical simulations of frictional manipulations
for GNRs up to $30$\,nm in length, one end of which is pushed or pulled
horizontally while held at different heights above the Au surface. 
The emergence of stick-slip originating from the short 1D edges rather
than the 2D ``bulk'', the role of adhesion, of lifting, and of graphene
bending elasticity in determining the GNR sliding friction are clarified
theoretically. 
The understanding obtained in this simple context is of additional value
for more general cases.
\end{abstract}


\maketitle

\section{Introduction}\label{sec.intro}

\begin{figure}
  \centering
  ~~~~~~~~~~\includegraphics[width = 0.8\columnwidth]{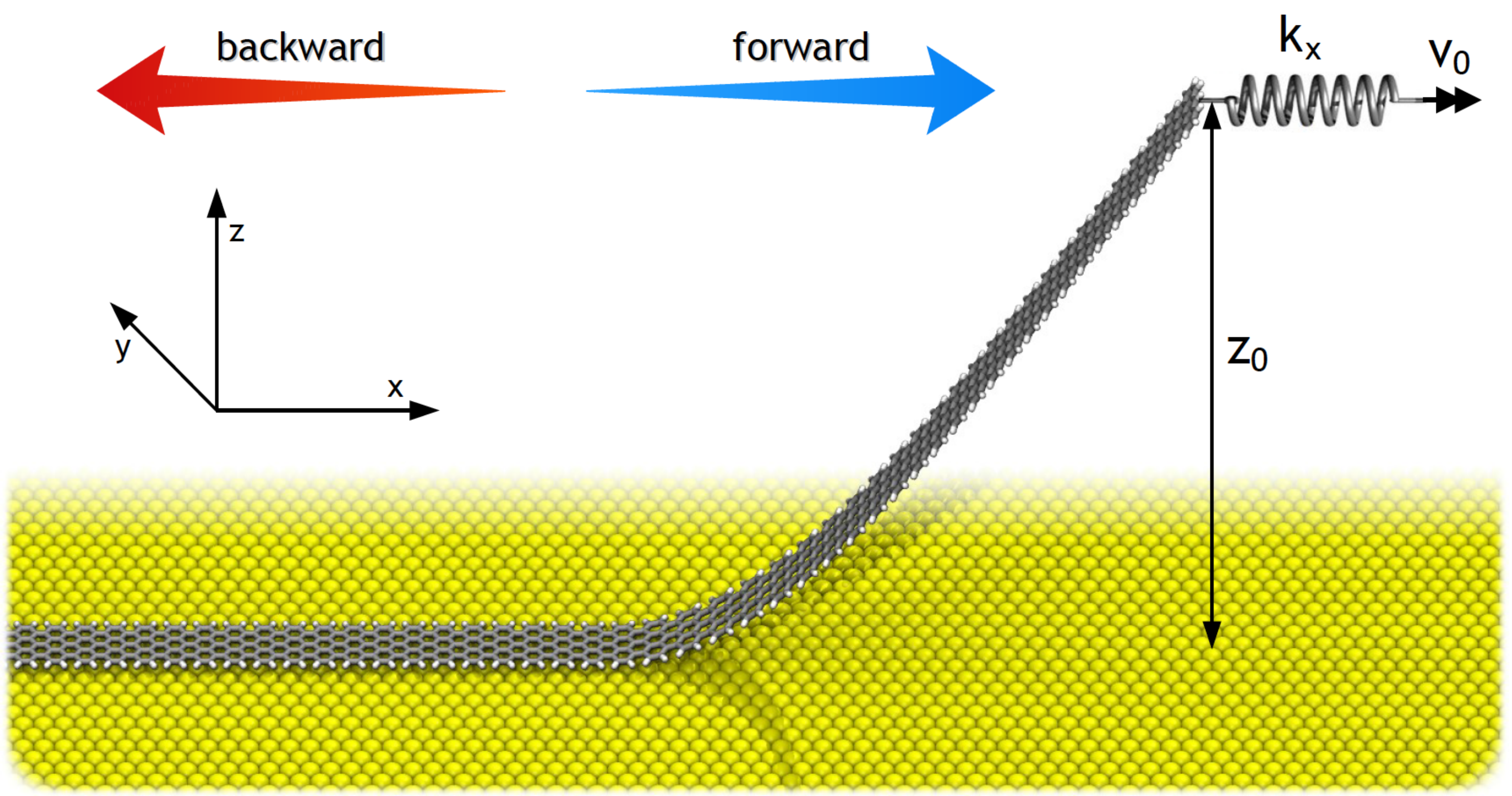}
  \caption{
    Schematic description of the setup used to simulate the AFM tip
    lifting the GNR at one end and pulling it laterally.
    One side of a soft spring is attached to the lifted end of the GNR,
    the other side is moving at constant positive or negative velocity,
    thus dragging the GNR forward or backward.
    The height $z_0$ of the lifted end is kept fixed in the simulations (see Method).
  }
  \label{setup}
\end{figure}

Nanofriction, a property of moving nanometer-sized interfaces widely 
investigated experimentally by atomic force microscopy (AFM), 
is progressively unveiling the detailed mechanisms which affect
the mechanical energy dissipation in well-controlled frictional setups 
\cite{Krim12,VanossiRMP13,Park14,Manini16,Manini17}.
Graphene is an important actor in this quest, because its strong
resilient structure makes it possible to push and slide flakes and planes once deposited on
suitable well-defined surfaces \cite{Dienwiebel04}.
Graphene nanoribbons (GNRs) too can be created and physisorbed on Au(111)
surfaces, by means of clever in-situ molecular-assembly techniques
\cite{cai10, Wang17}.
Once there, they can be picked up at one extreme and forced to slide by a
moving tip \cite{Kawai16}.
The dynamics of the GNR once dragged forward and backward (calling forward
the pulling, backward the pushing, as sketched in Fig.~\ref{setup}) may show
distinct regimes of motion depending on the lifting height, $z_0$.
At small lifting heights ($z_0=1\textrm{--}3$\,nm) there is an almost
symmetric behavior between forward and backward scans, not unlike that
observed experimentally for the low-lifted GNR \cite{Kawai16}.
At larger heights ($z_0=4\textrm{--}5$\,nm), different stick-slip patterns
and periodicities emerge with a substantial asymmetry between the two.

The present theoretical study aims at understanding the main features of
frictional dissipation in these systems.
%
\begin{figure*}
\centering
\includegraphics[width = 1.0\textwidth]{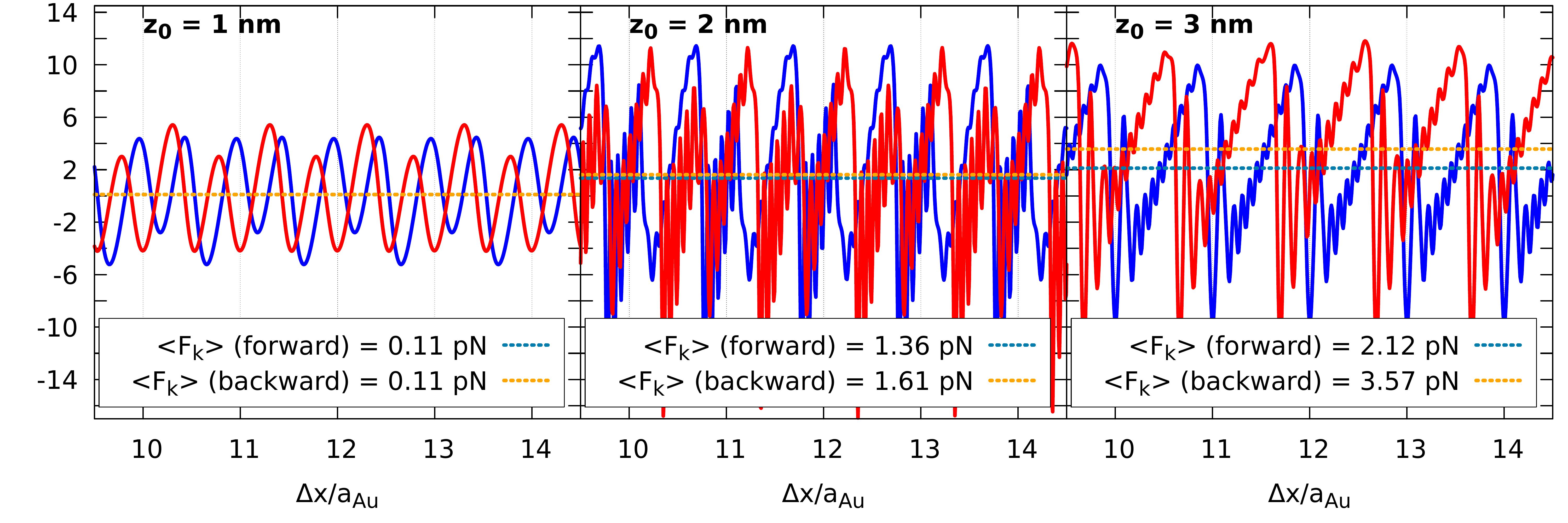}
\caption{
  Frictional force for GNR sliding at relatively small lifting heights
  $z_0=1\textrm{--}3$\,nm.
  The blue and red solid curves refer to the forward and backward sliding, respectively. 
  Dotted curves report the corresponding average frictional values.
}
\label{force-traces_1_2_3_nm}
\end{figure*}

\begin{figure}[!t]
\centering
\includegraphics[width = 0.50\textwidth]{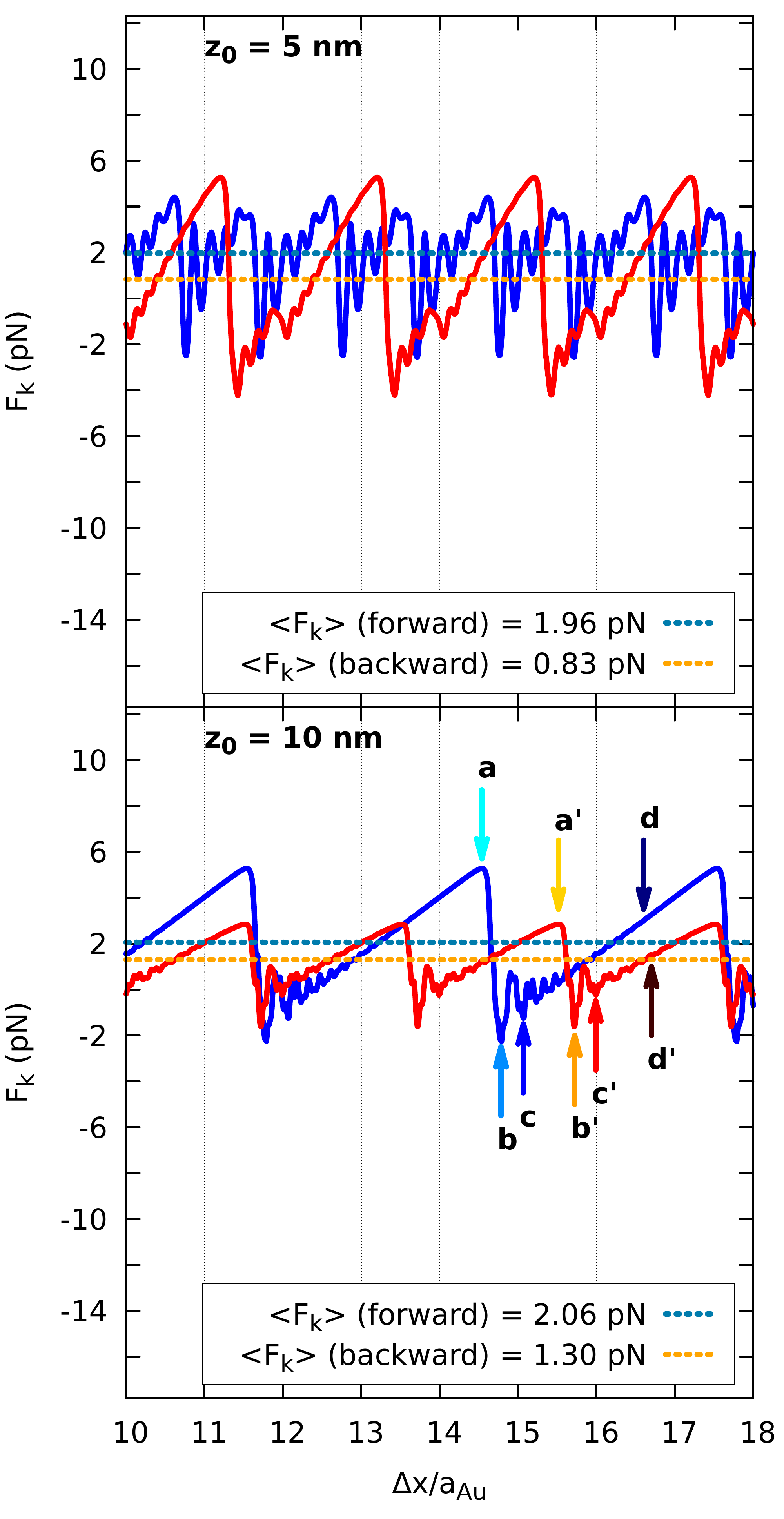}
\caption{
  Frictional force for GNR sliding at relatively large lifting heights
  $z_0=5$ and $10$\,nm.
  The blue and red solid curves refer to the forward and backward sliding, 
  respectively. Dotted curves report the corresponding average frictional values.
  Arrows identify some characteristic GNR configurations during the motion
  at $z_0=10$\,nm:
  (a) the end of the stick phase, (b) the end of the slip phase,
  (c) the beginning of stick, (d) half-stick.
  Non-primed and primed letters are for forward and backward motion,
  respectively.
  The GNR geometries for the configurations marked by arrows are reported
  in Fig.~\ref{z_profile_peel_pull} below.
}
\label{force-traces_5_10_nm}
\end{figure}

Anticipating our final conclusions, the forward-backward symmetric
frictional response at small lifting heights stems from the limited extent
of elastic deformations accumulated by the GNR when pulled against an energy barrier.
At increasing lifting height, the bending energy required to deform the GNR
decreases and the mechanical response under driving becomes different 
for the two opposite scan directions;
reaching the minimum energy needed to initiate sliding (the Peierls-Nabarro barrier \cite{Floria96}),
the GNR dynamics starts to develop asymmetric features in the emerging stick-slip regime 
for forward and backward pulling.
The main effects of this enhanced elastic deformation are an increased
period of the stick-slip motion and the occurrence of a possible ``peeling'' effect 
in the backward trace for increasing lifting height.

In addition, we show that the peaks of the time-resolved frictional force
traces depend critically on the effective contact length of the GNR section
still adhering to the substrate.
The force peak amplitudes exhibit an oscillation versus effective length of the GNR
mostly due to the imperfect compensation of the moir\'e superstructure
at the two ends of the physisorbed part of the GNR.
This effect is also related to the oscillatory behavior of the static
friction force versus size reported in the past for totally adhering
GNRs \cite{Gigli17}.

\section{System and method}\label{sec.system}

We simulate an armchair GNR, consisting of a stripe of alternating triplets
and pairs of carbon hexagons, of width $\sim0.7$\,nm and length
$\sim30.2$\,nm.
This length, a factor $\simeq 5$ larger than  that of our reference experiment
\cite{Kawai16}, enables us at the same time to reproduce qualitatively the
behavior of the force traces at small lifting height obtained
experimentally with a much shorter GNR, and to anticipate phenomena that
should come into play when the lifting height is sufficiently large
($z_0>5$\,nm), a regime where GNRs will undergo important elastic
deformations.
All the edge C-atoms at the periphery of the GNR are passivated with
hydrogens, in order to faithfully reproduce the experimental conditions
\cite{Kawai16}, and to obtain realistic peripheral C-C bond lengths, 
which are sensitive to saturation effects.

The simulated GNR is deposited on an unreconstructed Au(111) surface along the R30
direction, i.e.\ the GNR long axis lies parallel to the Au$[-1,0,1]$
crystallographic direction \cite{Gigli17}.
The atomistic dynamics of the GNR is simulated using the LAMMPS package
\cite{lammps} by means of a REBO force field \cite{Brenner02} for C-C and
C-H interaction, plus 2-body C-Au and H-Au interactions of the (6-12)
Lennard-Jones (LJ) type, as parametrized in Ref.~\cite{Gigli17}.
In the following we refer to these energy contributions as $V_{\rm REBO}$
and $V_{\rm LJ}$, respectively.

Starting from a fully relaxed GNR configuration, we lift progressively one
end row (three C atoms) of the GNR through a fictitious ultra-hard spring
($k_z= 1.6\cdot10^5$\,N/m), producing unilateral detachment up to a desired
height $z_0=1\textrm{--}13$\,nm (with respect to the unlifted GNR configuration), 
followed by a further relaxation in the lifted geometry.
After that, the mean coordinate of the lifted end of the GNR, while held
all the time at its fixed height $z_0$, is connected to a soft horizontal
pulling spring ($k_x = 1.5$\,N/m) and dragged forward or backward with
constant velocity $v_0 = \pm 0.5$\,m/s.
This procedure aims at mimicking, at least qualitatively, 
the lateral manipulation of a GNR, as done in AFM experiments \cite{Kawai16}.
While the real-time evolution of the underlying Au substrate is not
explicitly simulated, the GNR C and H atoms obey a dissipative Langevin
dynamics, at zero temperature and damping parameter $\gamma =
0.01$\,ps$^{-1}$, which prevents the externally-driven nanoribbon from
heating up.

The specific adopted $\gamma$ value ensures, we checked, a realistic
relative balance of inertial and dissipative terms, as discussed later, and does not
significantly affect the qualitative outcome of the simulated tribological
response within a quite broad range of values.

The equation of motion for each of the three C atoms of the lifted edge reads:
\begin{equation} \label{end_motion}
  m_{\rm C} \ddot{\mathbf{r}}_i
  = -  m_{\rm C} \gamma \dot{\mathbf{r}}_i
  - k_z(z_i - z_0) \hat{\mathbf{z}} - k_x(x_i - v_0 t) \hat{\mathbf{x}}
  - \nabla_{\mathbf{r}_i} V( \mathbf{r}_i, \{ \mathbf{R}_{\mu'} \} )
  \,,
\end{equation}
where $\mathbf{r}_i = (x_i, y_i, z_i)$ $(i = 1, 2, 3)$ are the positions of
the three C-atoms of the lifted edge, $\hat{\mathbf{x}}$ and
$\hat{\mathbf{z}}$ the unit vectors directed along the $x$- and $z$-axis,
and $V( \mathbf{r}_i, \{\mathbf{R}_{\mu'} \} ) = V_{\rm REBO}(\mathbf{r}_i,
\{\mathbf{R}_{\mu'} \}) + V_{\rm LJ}(\mathbf{r}_i, \{\mathbf{R}_{\mu'} \})$
is the total potential energy including the interaction among all GNR
particles and between particles and substrate.
The equation of motion for all the other atoms with coordinates
$\mathbf{R}_{\mu}$ is
\begin{equation} \label{gen_motion}
  m_{\mu} \ddot{\mathbf{R}}_{\mu}
  = -m_{\mu} \gamma \dot{\mathbf{R}}_{\mu}
  - \mathbf{\nabla}_{\mathbf{R}_{\mu}} V(\mathbf{r}_i, \{\mathbf{R}_{\mu'} \})
  \,.
\end{equation}

\section{Results and discussion}\label{results:sec}

We extract the instantaneous simulated frictional force as the elastic
force that the soft pulling spring exerts on the GNR
\begin{equation}
  F_{\rm k}(t) = 3 k_x \left[ v_0 t - x_{end}(t) \right]
\end{equation}
where $x_{end}(t) = \sum_{i = 1}^3\,x_i(t)/3$ is the mean $x$-coordinate of
the lifted end of the GNR, obtained by averaging the coordinates $x_i(t)$
of the three lifted-edge C atoms.

For each given lifting height $z_0$, the simulated AFM force trace is a plot
of $F_{\rm k}(t)$ as a function of 
time, or equivalently of the displacement of the fixed-speed end of the
spring $\Delta x(t) = |v_0| t$.
For ease of comparison, we express this displacement in units of the
lattice spacing of the gold substrate in the pulling direction,
$a_{\rm Au} = 2.8838$\,\AA.

Discarding initial transients, Figures \ref{force-traces_1_2_3_nm} and
\ref{force-traces_5_10_nm} show the steady-state simulated frictional
forces for lifting heights $z_0=1-3$\,nm, and $5-10$\,nm, respectively.
We note that, for a direct comparison highlighting intrinsically different features between the
forward (blue solid curves) and backward (red dashed curves) traces we show the latter 
forces reversed in sign and plotted versus positive displacements, thus not
displaying the typical dissipation frictional loop obtained for standard AFM 
back-and-forth scans.
At low lifting heights, $z_0 = 1-3$\,nm, Fig.~\ref{force-traces_1_2_3_nm}, 
the computed force traces for the forward and backward scans exhibit a
symmetric response, as observed in experiment \cite{Kawai16}.
Note that the experimental traces also contain a long-wavelength modulation
\cite{Kawai16} due to the Au-substrate reconstruction, here neglected.
At the small lifting of $z_0 = 1$\,nm the sliding force oscillation
reflecting the atomic corrugation on Au(111) is smooth in both
directions (see Movies in Supplementary data).
As a result, the average frictional force ($0.1$\,pN) is close to zero,
confirming the superlubric characteristic of the interface.
We note that, due to the lattice mismatch between the GNR structure and the underneath substrate 
along the considered R30 direction, there exist two inequivalent good matching interface
configurations, shifted almost by one half of the Au lattice spacing, giving rise approximately to 
a period doubling in the force traces. 

The frictional evolution for increasing lifting height is remarkable.
A first change in the dynamical response appears between $1$ and $2$\,nm
lifting.
At $z_0 = 2$\,nm the smooth sliding is replaced by atomic stick-slip 
with the same periodicity of the smooth oscillations at $1$\,nm.
With the occurrence of this intermittent dynamics, usually
marking in tribological systems the demise of superlubricity \cite{VanossiRMP13}, 
we very reasonably find that friction rises by an order of magnitude.
It can be noted that at the end of each slip the instantaneous force
oscillates considerably, in both forward and backward traces, due to 
inertial overshooting.
At higher lifting height $z_0 = 3$ a similar atomic stick slip is again observed,
but without the delicate superimposed period duplication observed at smaller  $z_0$.

A different scenario emerges for higher lifting, such as $z_0 = 5$ and
$10$\,nm, Fig.~\ref{force-traces_5_10_nm}.
Forward and backward traces are not symmetric anymore, and multiple jumps
\cite{Medyanik06} start to show up (see Movies in Supplementary data), 
contrasting the basically single stick-slip regime observed at small lifting.
The slip distance depends quite generally on the lifting height, which
controls the mechanical softness of the lifted part, and on the pulling
direction.
For instance, at $z_0=5$\,nm the forward force trace is single slip, while
that of the corresponding backward scan becomes double.
Conversely, at $z_0=10$\,nm the forward trace shows a stick-slip period of
three lattice spacings, as opposed to two lattice spacings in the backward
case.

Such asymmetric response, as we shall see, is determined by the interplay
of two main effects.
Firstly, forward and backward configurations imply different
effective contact areas between the GNR and the substrate.
Since the static friction oscillates widely with GNR contact length $L_c$
\cite{Gigli17}, small differences in the effective contact length can lead
in general to quite different static-friction thresholds.
As a result, small differences in $z_0$ may lead to different
dynamical friction patterns.

Secondly, as detailed in Sect.~\ref{sec:energy} below, the interplay
between bending energy and adhesion differs strongly in the two pulling
directions.

\begin{figure}[!t]
  \centering
  \includegraphics[width = 0.50\textwidth]{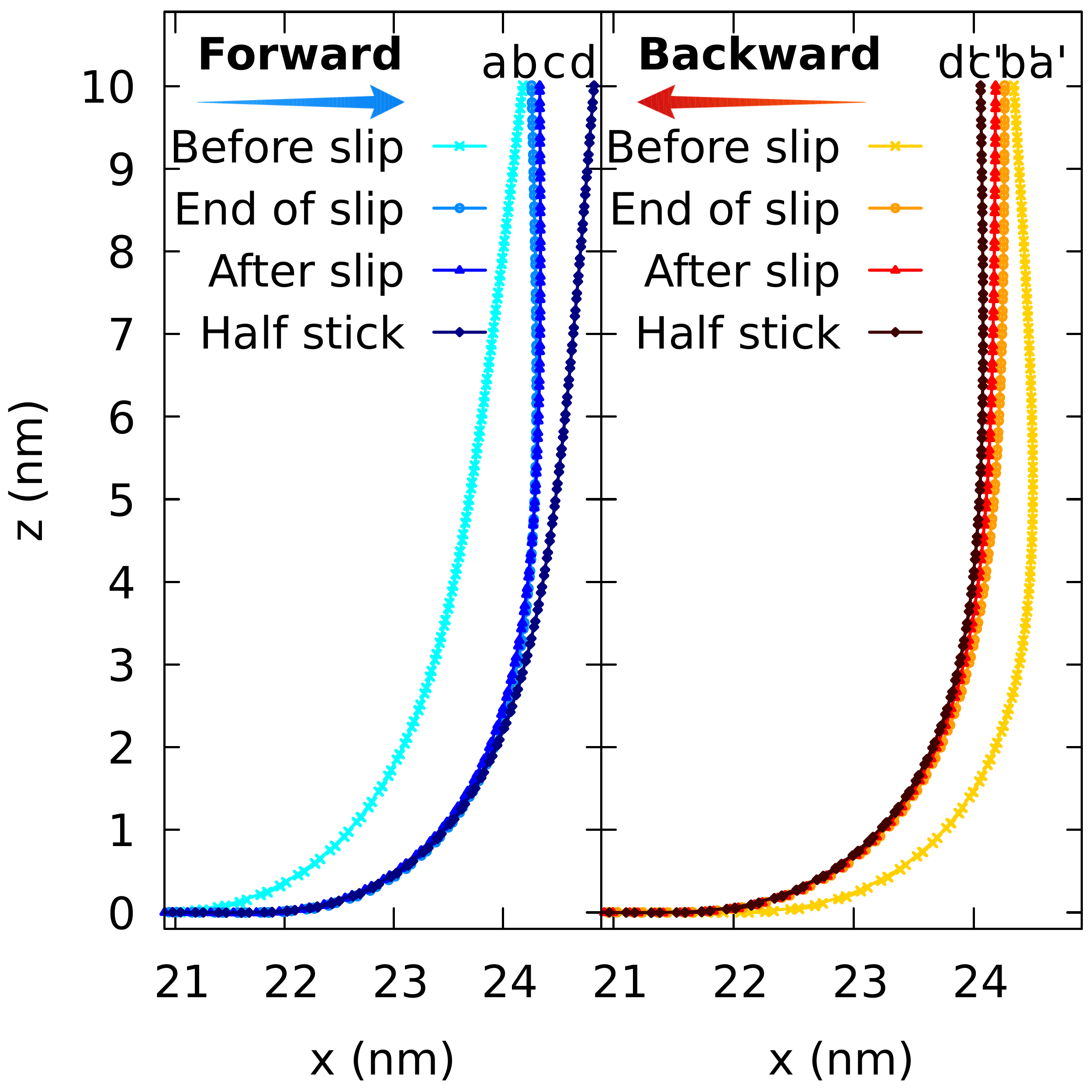}
  \caption{
    The GNR lateral profile at $z_0=10$\,nm during the forward (left panel)
    and backward (right panel) motion for the four successive configurations
    marked by arrows in Fig.\ \ref{force-traces_5_10_nm}.
  }
  \label{z_profile_peel_pull}
\end{figure}

A first insight in the different forward and backward GNR dynamics can be
obtained by examining the characteristic shape of the GNR at specific
instants during the stick-slip motion.
Figure~\ref{z_profile_peel_pull} shows the lateral profile of the GNR
in the forward and backward motion at $z_0=10$\,nm, at four distinct
instants marked by arrows in Fig.~\ref{force-traces_5_10_nm}.
The main features of the stick-slip dynamics in the forward and backward motion are very similar:
once the spring reaches the critical elongation to overcome the Peierls-Nabarro barrier (a/a'), 
a slip event occurs, the physisorbed section sprints forward/backwards and reaches 
a new pinned position (b/b'), with a new {\em increase} (in the forward motion) /{\em decrease} (in the
backward motion) of the GNR bending energy,
which is then progressively {\em released}/{\em absorbed}
 during the subsequent stick phase (c$\rightarrow$d / c'$\rightarrow$d'). 

\section{Energy considerations}
\label{sec:energy}

\begin{figure}[h!]
  \centering
  \includegraphics[width = 0.50\textwidth]{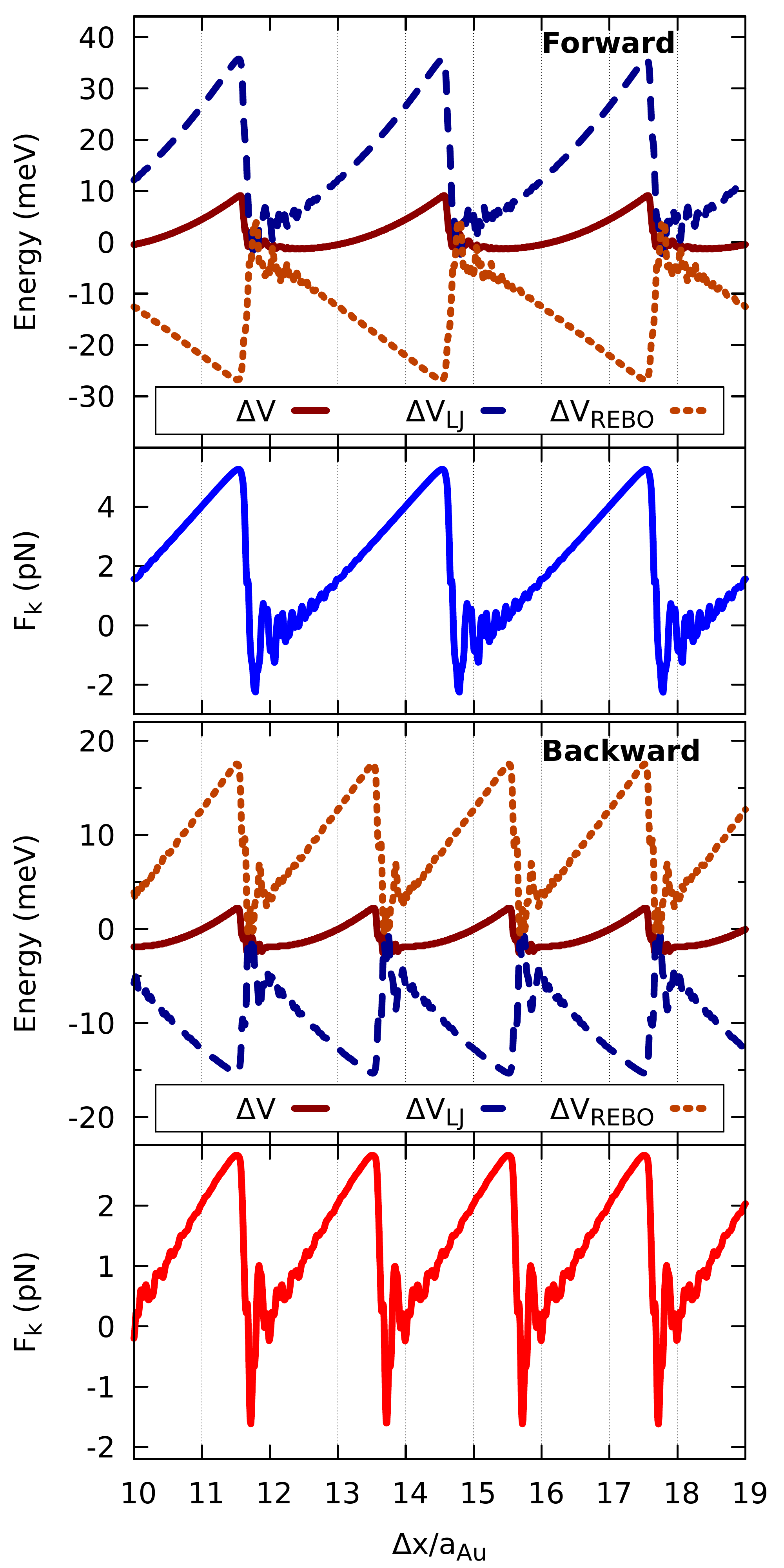}\\
  \caption{
    Comparison between the frictional force and the variation of the
    elastic intra-ribbon $V_{\rm REBO}$ and adhesive ribbon-substrate
    $V_{\rm LJ}$ contributions to the GNR total energy in the stick-slip
    motion at $z_0=10$\,nm.
  }
  \label{energy_contributions}
\end{figure}

It is instructive to analyse how the individual energy contributions
coming from the elastic bending of the GNR and from the adhesion to the
substrate evolve during the stick-slip frictional dynamics.
Consider for instance the motion of the GNR at large $z_0=10$\,nm.
The  total GNR potential energy $V$ is the sum of an intra-ribbon term,
$V_{\rm REBO}$ from the C-C and C-H bonds, which controls the  
planar and bending stiffness, plus a second term, $V_{\rm LJ}$, stemming from
the C-Au and H-Au interactions which controls the adhesion of the
unlifted part of the GNR.
The time variation of $V$ with respect to our reference configuration at
$t = 0$, (a relaxed GNR with one lifted end), can be written as
\begin{equation}\label{energy_var}
  \Delta V(t) = \Delta V_{\rm REBO}(t) + \Delta V_{\rm LJ}(t)
  \,.
\end{equation}

\begin{figure}[h!]
  \centering
  \includegraphics[width = 0.50\textwidth]{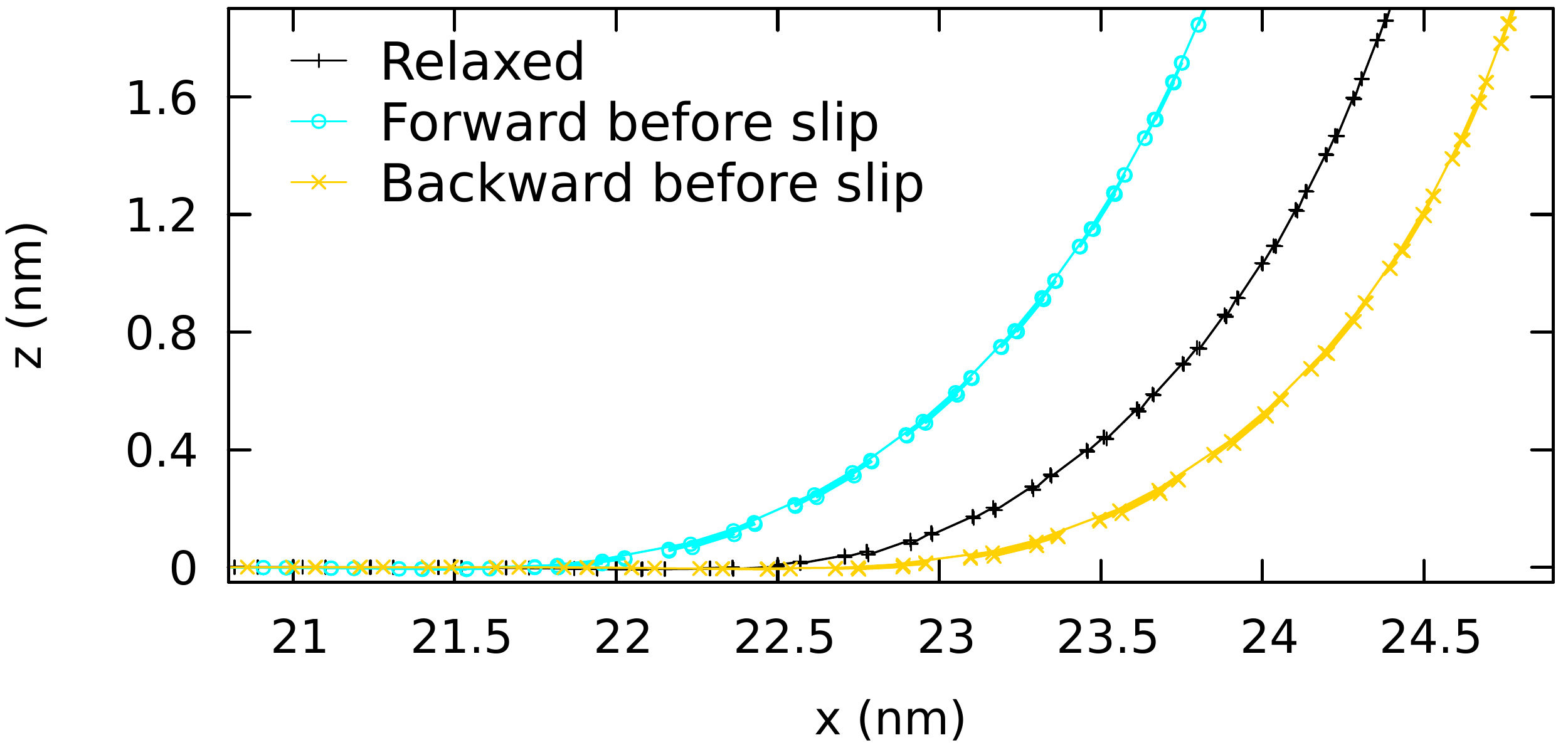}
  \caption{
    Lateral profile of the GNR in the forward (cyan line) and backward
    motion (yellow line) just before the slip.
    The configuration of the relaxed GNR at rest (black line) is included
    as reference.
  }
  \label{Contact_10nm_pull_vs_peel}
\end{figure}

For forward and backward motion, Fig.~\ref{energy_contributions}
compares the frictional force evolution (already displayed in
Fig.~\ref{force-traces_5_10_nm}) and that of the potential energy terms
$\Delta V_{\rm REBO}$, $\Delta V_{\rm LJ}$ and $\Delta V$.
Note the opposite contributions to the total GNR energy for
forward and backward sliding.
In the forward scan, the intra-ribbon contribution $\Delta V_{\rm REBO}$
is negative, with an energy gain due to the decrease of GNR curvature in
the detached part, as discussed in Sect.~\ref{results:sec} above.
At the same time, the system loses adhesive energy, not just because the
external force works to overcome the static friction energy barrier which blocks 
the sliding, but mainly because the physisorbed section shortens in length as
the GNR end is pulled forward (see also the zoomed-in GNR $z$-profile in
Fig.~\ref{Contact_10nm_pull_vs_peel}), causing an increase of $\Delta
V_{\rm LJ}$.
Exactly the opposite occurs for backward sliding, where $\Delta V_{\rm
  REBO}$ is positive, owing to the curvature increase of the detached part,
whereas $\Delta V_{\rm LJ}$ is negative reflecting a corresponding
improvement of adhesion due to an increased contact length $L_c$ (see again
Fig.~\ref{Contact_10nm_pull_vs_peel}).
For completeness, we note that, at even larger $z_0$ values, the
backward-driven GNR may initiate to peel off the Au surface during the stick phase, 
thus starting decreasing the $\Delta V_{\rm LJ}$ adhesive contribution.

\begin{figure}[!t]
  \centering
  \includegraphics[width = 0.40\textwidth, height = 0.85\textheight]{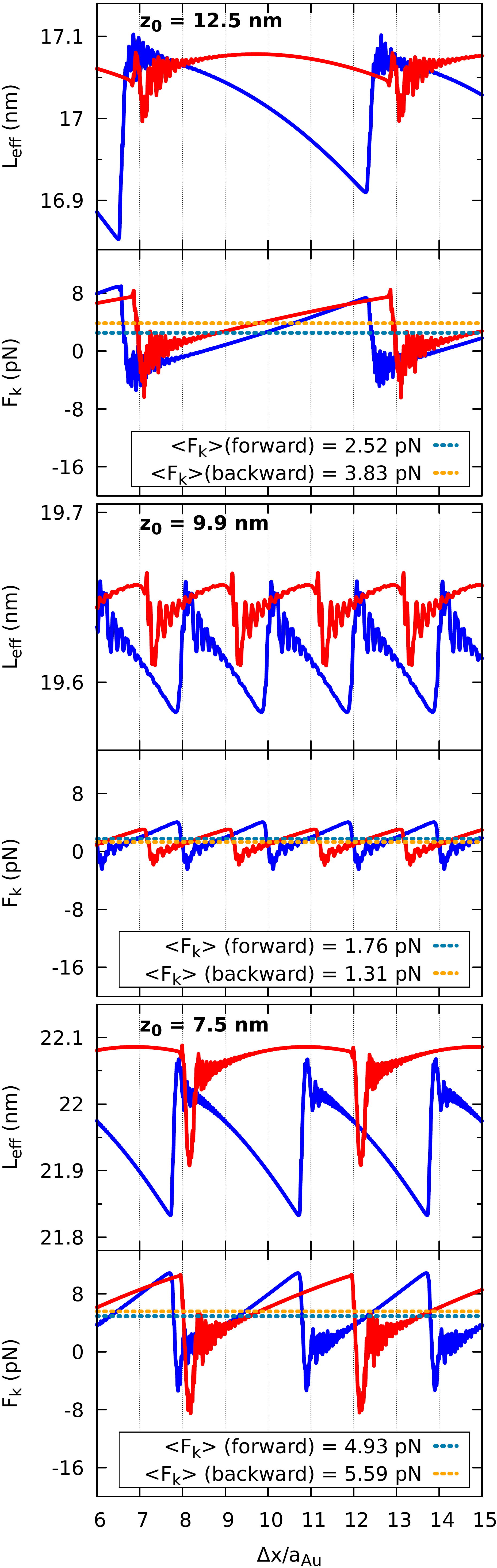}
  \caption{
    Comparison between the friction-force traces for three large
    lifting heights, $z_0=7.5$, $9.9$ and $12.5$\,nm, and the change of the
    effective instantaneous contact length $L_{\rm eff}$
    between the GNR and the substrate.
    As previously, blue and red solid curves refer to forward and backward sliding, respectively, 
    and dotted curves report the corresponding average values.
  } 
  \label{force-traces_7_10_13_nm}
\end{figure}

There is a clear correspondence between the general energy evolution
described above and the lifted nanoribbon geometry.
Figure~\ref{Contact_10nm_pull_vs_peel} compares the shape profile
$z=z(x)$ of the GNR near the detachment point, just before the slip
either forward or backward.
By comparison with the relaxed, static shape (zero force), the curvature and 
the physisorbed section of the GNR are respectively smaller in the forward case, 
and larger in the backward case.

\section{Role of the ribbon short edge and uncompensated moir\'e pattern}\label{sec.uncompensation}

As was observed in our previous study of the fully adhering -- non-lifted
-- GNRs \cite{Gigli17}, the 2D ``bulk'' of the GNR/Au(111) interface is
incommensurate and structurally lubric (``superlubric'').
Like in other superlubric systems, the static friction --
the minimal force required to set the interface into sliding motion -- does
not grow (on average) as much as the contact area.
Specifically, for a non-lifted GNR, the static friction oscillates around 
a fairly constant mean value as a function of the nanoribbon length \cite{Gigli17}.
This indicates that the edges, here the short ones, are mostly responsible for pinning -- a
feature similarly found in incommensurate rare-gas islands deposited on metal surfaces \cite{Varini15}
and in twisted bilayer graphene \cite{Koren16b}.
The strong oscillation of the static friction $F_{\rm s}$ around the constant average trend 
as a function of the GNR length is related to the ``uncompensated'' moir\'e pattern near the GNR ends, 
i.e.\ the residual of $L_c$ divided by the moir\'e-pattern wavelength.
This friction oscillation may involve variations in $F_{\rm s}$ comparable with 
the average \cite{Koren16b}. This appears to be the case also with
lifted GNRs, where the effective contact length $L_{\rm eff}$, defined below, 
varies as a function of $z_0$ and changes dynamically in time.

By lifting the GNR at successively increasing heights, the effective
contact length $L_{\rm eff}$ will change, giving rise to minima/maxima of the static
friction force. 
We define the effective contact length $L_{\rm eff}$ of a lifted GNR by
dividing $V_{\rm LJ}$ by the same quantity per unit length of an
infinite-length simulated GNR with periodic boundary conditions, $V_{\rm
  LJ}^{\infty}$:
\begin{equation}\label{effective_length}
  L_{\rm eff}(z_0,t) = \frac{V_{\rm LJ}(z_0,t)}{V_{\rm LJ}^{\infty}}
  \,,
\end{equation}
where $V_{\rm LJ}(z_0,t)$ is the total interaction energy between the
GNR and the substrate at the lifting height $z_0$ and at time $t$.
It turns out that for lifting heights between $z_0=7.5$\,nm and
$z_0=12.5$\,nm we cover one complete period of the static friction \cite{Gigli17}.
Figure~\ref{force-traces_7_10_13_nm} shows the force traces corresponding
to lifting heights $z_0=7.5$, $9.9$, and $12.5$\,nm, the first and the last
ones corresponding to expected local maxima of the oscillating static
friction versus effective size, the second to a local minimum, along with
the corresponding change in time of the effective contact length $L_{\rm
  eff}(z_0,t)$ of Eq.~(\ref{effective_length}).
As expected, the peaks in $F_{\rm k}$ and the mean friction forces are
larger at lifting heights that correspond to the expected local maxima of
static friction, namely $z_0=7.5$\,nm and $z_0=12.5$\,nm, than at the
expected local minimum, namely $z_0=9.9$\,nm.

\begin{figure}[!t]
\centering
\includegraphics[width = 0.50\textwidth]{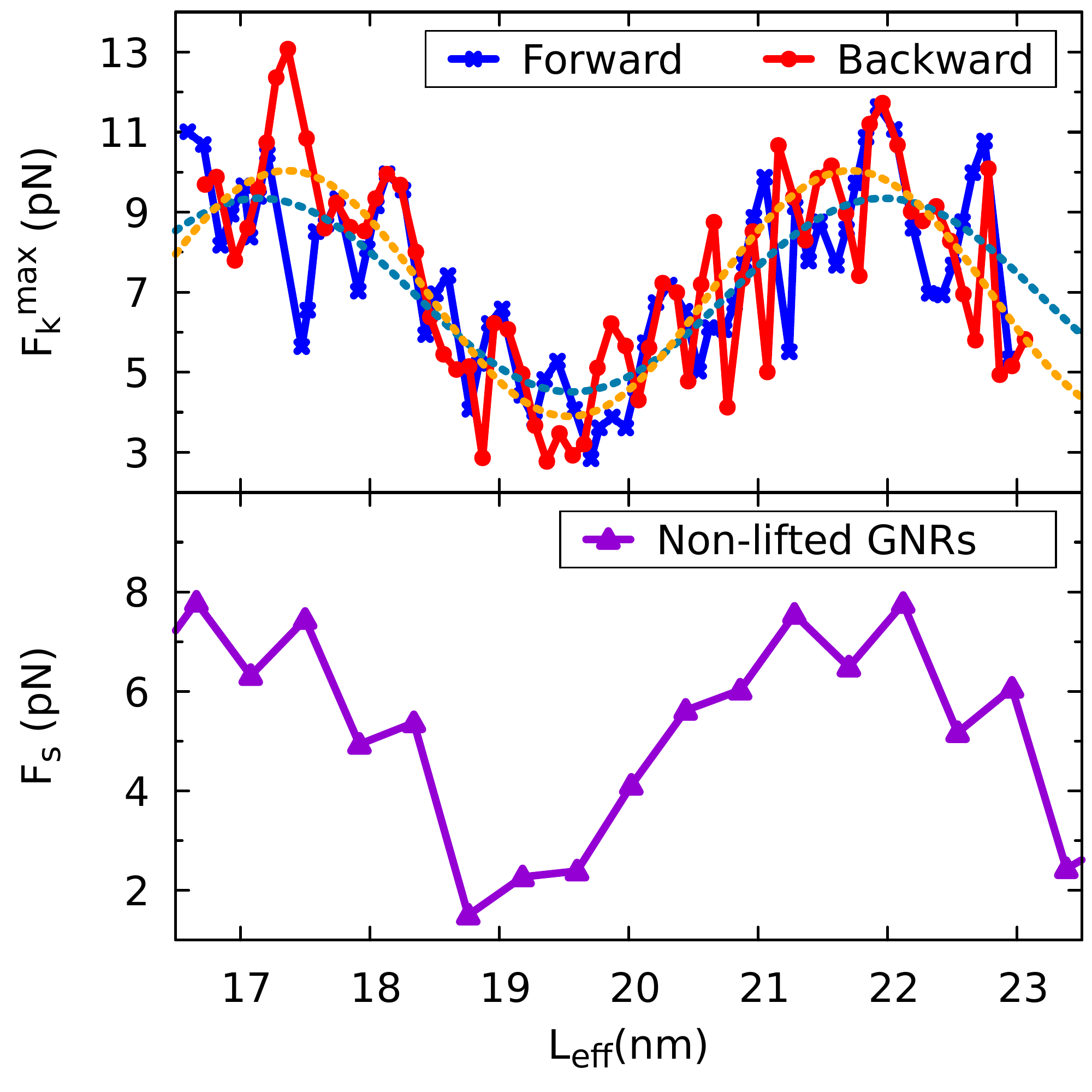}
\caption{
  Comparison between the static friction force of non-lifted GNRs
  \cite{Gigli17} (lower panel), and the maximum kinetic force (upper
  panel), as a function of the effective contact length $L_{\rm eff}$,
  varied by repeating the simulations for many lifting heights $z_0$.
  The dotted cyan and orange lines are the corresponding best fitting
  curves of Eq.~(\ref{fitting_functions}) to the $F_{\rm k}^{\rm max}$ data.
}
\label{Forcemax_vs_effective_length}
\end{figure}

For a grid of lifting heights $z_0$, Fig.~\ref{Forcemax_vs_effective_length}
reports the maximum force $F_{\rm k}^{\rm max}$ obtained from the peaks just 
before slip once a steady stick-slip regime is established versus the
effective contact length $L_{\rm eff}$ for that height.
The best fitting sinusoids of the form
\begin{equation}\label{fitting_functions}
  F_{\rm k}^{\rm max}(L) = \alpha + \beta \sin \left(\frac{2 \pi L}{\lambda_m} - \delta \right),
\end{equation}
for both the forward and backward motion, are also drawn as reference.
$\alpha$, $\beta$, $\lambda_m$, $\delta$ are fitting parameters.
The values and the $\lambda_m = 4.86$\,nm period oscillation of the lifting-dependent
maximum force compares reasonably well with the established static friction
trend as a function of the non-lifted GNR length \cite{Gigli17}.
Somewhat larger in magnitude, both forward and backward maximum forces
share the same oscillation as the static friction of the non-lifted GNR
with length equal to the effective lifted GNR length $L_{\rm eff}$.
This result further confirms that the uncompensated, edge-related, part of
the moir\'e pattern determines the magnitude of the maximum
kinetic-friction force before slip.

Figure~\ref{Forcemax_vs_effective_length} compares the maximum kinetic
friction $F_{\rm k}^{\rm max}$ with the static friction
$F_{\rm s}$ obtained for fully adhering GNRs \cite{Gigli17}.
For a given system in the underdamped regime, the two quantities should
match in the limit of vanishing driving velocity $v_0$.
At finite velocity it is generally expected that $F_{\rm s} > F_{\rm k}^{\rm max}$, 
with static friction always exceeding dynamic one.
Here, naively, we observe the opposite. 
This might look counterintuitive, as one might expect a larger friction for fully adhering GNRs.
However, as pointed out above, static friction is dominated by the two GNR short-ends in this 
superlubric system.
The two short-edges are equivalent in the unlifted case, and are responsible 
for the frictional oscillations as a function of $L_c$ \cite{Gigli17}.
By contrast, in the case of lifted GNR the bending at the leading edge
produces a termination which is strongly inequivalent to that of the
trailing edge.
As a result, cancellation of the lateral forces acting on the two ends is
more problematic, yielding generally an overall larger friction.

It is also worth asking if GNRs might show any tendency to peel off
the substrate when driven backward at large lifting heights. 
%
As seen in Figures \ref{energy_contributions} and \ref{Contact_10nm_pull_vs_peel}, 
the backward stick-slip motion is accompanied by an {\em increase} of adhesion 
in the stick state, while a {\em decrease} of adhesion is seen in the forward motion.
In Figure \ref{force-traces_7_10_13_nm}, $z_0=9.9$\,nm, this fact is
confirmed by the increase of $L_{\rm eff}$
in the stick state of the backward motion.
In these cases no tendency to peel off is registered.
In contrast, at a lifting height of $12.5$\,nm, we notice that 
in the backward motion the adhesive length increases up to a maximum and then
decreases again with a sort of parabolic trend.
This indicates that the spring initially pushes the physisorbed atoms
adjacent to the bent GNR section down in closer contact with substrate,
promoting an increased adhesion.
Once the extension of the driving spring is sufficiently large, the GNR
starts to detach from the substrate, causing a loss of adhesion.
This analysis shows that, depending on the lifting height $z_0$ and the
precise value of the static friction barrier at that height, the GNR can
indeed start to peel off from the substrate.
In all simulated cases, as backward pulling continued, a slip event would
release the bending stress before the peeling instability would fully
develop and lead the GNR to a complete peel off.
As a general rule, peeling is more pronounced for those combinations of
$z_0$ and GNR length leading to those $L_{\rm eff}$ producing the largest
possible static friction threshold, and generally for larger lifting
height, because of the softer GNR elasticity and greater mechanical
advantage.

\section{Conclusions}

We have investigated the dynamical friction of lifted graphene nanoribbons
on a Au(111) substrate by means of non-equilibrium molecular-dynamics
simulations.
Mimicking the experimental setup of Kawai et al.\ \cite{Kawai16} we
reproduce and interpret the observed frictional regimes of the GNR as a function of the lifting height $z_0$.
For increasing $z_0$, we predict a remarkable transition from smooth
sliding to atomic stick-slip, characterized initially by single slips, and then by
multiple slips at larger heights.
Specifically, the periodicity of the stick-slip dynamics is dominated by
the bending elasticity of the GNR, which enables larger slip distances at
larger heights.
The augmented softness, introduced by bending of the GNR as $z_0$ increases, 
plays opposite roles for the two driving directions, decreasing (forward) and increasing (backward) 
the GNR/substrate adhesion.   
The lifting-dependent amplitude of the instantaneous friction force is not a "bulk" feature, 
and is entirely determined by the short edges of the GNR --
in the lifted case as well as in the non-lifted case.

We find an oscillation of friction with lifting height. That in turn is related, via identification 
of an effective GNR contact length of the physisorbed GNR section, to the moir\'e-pattern 
lack of compensation close to the edges, qualitatively similar but quantitatively different 
to that occurring in the static friction of unlifted GNRs
\cite{Gigli17}.
Past experiments on lifted GNR sliding \cite{Kawai16}
have not yet explored the new regime which we describe here, essentially due to 
the relatively small length of the GNR used there ($6.28$\,nm only), whereby 
the GNR lifted at $5$\,nm was almost completely detached from the Au-substrate, 
very nearly peeled off. 
Our much longer -- 30\,nm -- simulated GNR,  
only approaches peeling at lifting heights larger than $\sim10$\,nm,
as shown by the time evolution of the effective contact length.
Present predictions about the sliding should be borne out by future experiments, hopefully on longer GNRs,
as well as on more general physisorbed flakes of graphene and other 2D materials. 
 
In these systems, they should  be able to find, for increasing lifting heigthts,  
a transition from smooth sliding to stick-slip, the asymmetric forward/backward friction, 
and a peel-off instability. 

\vspace*{0.5 cm}

\subsection*{Acknowledgments}

\noindent We acknowledge useful discussions with E.\ Meyer, E.\ Gnecco and A. Benassi.
\noindent Work in Trieste was carried out under ERC Grant 320796 MODPHYSFRICT.
\noindent The COST Action MP1303 is also gratefully acknowledged.

{\small
\section*{Acknowledgments}
We acknowledge useful discussions with E.\ Meyer, E.\ Gnecco and A. Benassi.
This work was supported by the ERC Advanced Grant No. 320796-MODPHYSFRICT.
The COST Action MP1303 is also gratefully acknowledged.
}

\vspace{3mm}

\newcommand*{\xdash}[1][3em]{\rule[0.5ex]{#1}{0.55pt}}
\xdash[10em]

\bibliographystyle{iopart-num}
\bibliography{biblio}

\end{document}